\documentclass[12pt]{article}
\usepackage{epsfig}

\makeatletter
\newcommand{\eqref}[1]{(\ref{#1})}
\newcommand{\Chref}[1]{\expandafter\MakeUppercase\chaptername~\ref{#1}}
\newcommand{\Secref}[1]{\expandafter\MakeUppercase\secrefname~\ref{#1}}
\newcommand{\Appref}[1]{\expandafter\MakeUppercase\appendixname~\ref{#1}}
\newcommand{\Figref}[1]{\expandafter\MakeUppercase\figurename~\ref{#1}}
\newcommand{\Tbref}[1]{\expandafter\MakeUppercase\tablename~\ref{#1}}
   \let\secref\Secref  
\let\figref\Figref  
\newcommand{\secrefname}{Section}
\DeclareSymbolFont{AMSb}{U}{msb}{m}{n}
\DeclareSymbolFontAlphabet{\mathbb}{AMSb}
\newcommand{\inv}{^{\raise.15ex\hbox{${\scriptscriptstyle -}$}\kern-.05em 1}} % inverse
\newcommand{\vev}[1]{\langle #1 \rangle}                  % pairing
\newcommand{\e}{\mathrm{e}}                               % exp e
\newcommand{\dint}[2][]{\mathop{\mathalpha{\int#1}#2}}    % integral with density
\newcommand{\doint}[2][]{\mathop{\mathalpha{\oint#1}#2}}  % o-integral with density
\newcommand{\set}[1]{\mathbb{#1}}                         % font used for sets
\newcommand{\Z}{\set{Z}}
\newcommand{\group}[1]{\mathop{\kern\z@\mathrm{#1}}\nolimits}     % font used for groups
\newcommand{\U}{\group{U}}                                % groups
\newcommand{\SU}{\group{SU}}      
\newcommand{\SO}{\group{SO}}
\newcommand{\Sp}{\group{Sp}}
\newcommand{\USp}{\group{USp}}            
\newcommand{\opname}[1]{\mathop{\kern\z@\mathrm{#1}}\nolimits}    % font used for operators
\newcommand{\Tr}{\opname{Tr}}                             % Trace         
   \newcommand{\CD}{\mathcal{D}}
   \newcommand{\CF}{\mathcal{F}}

   \newcommand{\CN}{\mathcal{N}}
\newcommand{\CO}{\mathcal{O}}

\newcommand{\CW}{\mathcal{W}}   
   
\renewcommand{\section}{\@startsection{section}{1}{\z@}%
                                    {-7ex \@plus -1ex \@minus -.2ex}%
                                    {2.5ex \@plus.2ex}%
                                    {\normalfont\large\scshape\centering}}
\renewcommand{\subsection}{\@startsection{subsection}{2}{\z@}%
                                       {-5ex \@plus -1ex \@minus -.2ex}%
                                       {1.5ex \@plus.2ex}%
                                       {\normalfont\normalsize\scshape}}

\renewcommand\@seccntformat[1]{\ignorespaces\csname #1name\endcsname\space
                               \csname the#1\endcsname.\quad}   % Extra period and name added
\newdimen\captionmargin 
\newdimen\captionindent 
\newdimen\captionwidth 
\newcommand{\captionfont}{\slshape}
\newcommand\@captionlabel[1]{\textsc{#1:}\space}
\long\def\@makecaption#1#2{%
  \vskip\abovecaptionskip  
  \captionwidth\hsize 
  \advance\captionwidth -2\captionmargin
  \sbox\@tempboxa{\@captionlabel{#1}\captionfont #2}%
  \ifdim \wd\@tempboxa >\captionwidth
    \ifdim\captionindent>\z@ 
      \advance\captionwidth -\captionindent
      \hskip\captionindent
    \fi
    \hskip\captionmargin
    \parbox[t]{\captionwidth}{\leavevmode\hskip-\captionindent
      \@captionlabel{#1}\captionfont #2}%
  \else
    \global \@minipagefalse
    \hb@xt@\hsize{\hfil\box\@tempboxa\hfil}%
  \fi
  \vskip\belowcaptionskip}
\def\eqnarray{%
   \stepcounter{equation}%
   \def\@currentlabel{\p@equation\theequation}%
   \global\@eqnswtrue
   \m@th
   \global\@eqcnt\z@
   \tabskip\@centering
   \let\\\@eqncr
   $$\everycr{}\halign to\displaywidth\bgroup
       \hskip\@centering$\displaystyle\tabskip\z@skip{##}$\@eqnsel
      &\global\@eqcnt\@ne$\;\hfil{##}$\hfil
      &\global\@eqcnt\tw@$\;\displaystyle{##}$\hfil\tabskip\@centering
      &\global\@eqcnt\thr@@ \hb@xt@\z@\bgroup\hss##\egroup
         \tabskip\z@skip
      \cr}
\ifcase \@ptsize
\or 
\or 
\fi
  \divide\@tempdima\baselineskip
  \@tempcnta=\@tempdima
\makeatother
\begin{document}

%
%--------- Titlepage ---------------------------------------------------------
%

\thispagestyle{empty}

\begin{flushright}\scshape
RUNHETC-2002-50\\
hep-th/0212095\\
December 2002
\end{flushright}
\vskip1cm

\begin{center}

{\LARGE\scshape
Super Yang-Mills With Flavors From 
Large $N_f$ Matrix Models
\par}
\vskip15mm

\textsc{Christiaan Hofman$^{1\dagger}$}
\par\bigskip
{\itshape
  ${}^1$New High Energy Theory Center, Rutgers University,\\
        136 Frelinghuysen Road, Piscataway, NJ 08854, USA,}
\par\bigskip
\texttt{${}^\dagger$hofman@physics.rutgers.edu}

\end{center}
\vspace*{1cm}

\section*{Abstract}

We consider the exact effective superpotential of $\CN=1$ $\U(N_c)$ super Yang-Mills 
theory with $N_f$ massive flavors an additional adjoint Higgs field. 
We use the proposal of Dijkgraaf and Vafa to calculate the 
superpotential in terms of a matrix model with a large number of flavors. 
We do this by gauging the flavor symmetry and forcing this sector in a classical vacuum. 
This gives rise to a 2-matrix model of ADE type $A_2$, and large flavors. 
This approach allows us to add an arbitrary polynomial tree level superpotential 
for the Higgs field, and use strict large $N$ methods in the matrix model.

\newpage
\setcounter{page}{1}
%
%--------- Main Article ------------------------------------------------------
%

\section{Introduction}

Recently, Dijkgraaf and Vafa (DV) proposed an exact calculation of 
superpotentials in $\CN=1$ super Yang-Mills theories using matrix 
models \cite{dv1,dv2,dv3}. Later the derivations of the relation to 
the matrix models was shown directly from the field theory in 
\cite{dijgrilavaza,cadoseiwit}. 
The DV proposal has initiated much research, extending the proposal to include 
flavors, other gauge groups, gravitational corrections, and relation to 
Seiberg Witten curves and field theory calculations of the effective 
superpotential \cite{arcafehe1}-\cite{seki}. 

The basic proposal of DV considers an $\U(N_c)$ gauge theory with an adjoint Higgs field 
$\Phi$, which has a tree level superpotential $\Tr W(\Phi)$. The solution is given in 
terms of a matrix model with of a $M\times M$ matrix $\Phi$ with action 
$\frac{1}{g_s}\Tr W(\Phi)$. The effective superpotential for the glueball 
field $S=\vev{\lambda\lambda}$ can be calculated as the derivative of the 
planar contribution to the free energy in the matrix model 
\begin{equation}
  W_{eff}(S) = N_{c}\frac{\partial\CF_0}{\partial S}, 
\end{equation}
identifying $S$ with the 't Hooft coupling $g_sM$ of the matrix model. 
As only the planar diagrams contribute, this can be calculated by large $N$ 
techniques. 

In the approach to the addition of flavors, there are two proposals. 
First in \cite{arcafehe1} it is conjectures that one has to add the 
planar free energy with 1 quark-loop to the above effective action. 
This proposal has been used in most of the other 
approaches to the use of matrix models for super Yang-Mills with flavors, 
and gives agreement with previously known results. 
The proposal of \cite{mcgre} amounts to taking also the number of 
flavors $M_f$ in the matrix model to infinity, but holding the ratio 
$M_f/M$ fixed. One can easily convince oneselve that this gives 
contributions from all quark loops, so the proposals seem to be 
incompatible. One can argue that there should only contributions from 
planar diagrams with at most one quark loop to the effective superpotentials. 
This is a generalization of the arguments used in \cite{dv1,cadoseiwit} 
to show that the effective superpotential has only contributions from the planar 
diagrams and is linear in $N_c$. 
What makes the proposal of \cite{arcafehe1} harder to work with is the fact that 
one needs $1/N$ corrections. These could in principle be calculated using 
techniques developed in \cite{amchekrima,ake}. In most other papers on the 
subject only quadratic tree level superpotentials were considered, 
for which the planar diagrams can be summed explicitly. 

In this paper we will show that one can add flavors to DV staying purely in a large $N$ limit. 
As in \cite{mcgre} we consider a matrix model with a large number of flavors $M_f$, 
but we will not constrain their ratio. This would add an extra large $N$ parameter 
$S_f=g_sM_f$, which we will take to zero at the end of the calculation. 
The proposed formula for the effective superpotential is then 
\begin{equation}\label{weffprop}
  W_{eff}(S) = N_c\frac{\partial\CF_0}{\partial S}+N_f\frac{\partial\CF_0}{\partial S_f}.
\end{equation}
A simple counting of the dependence of the contributions on $S$ and $S_f$ 
shows that there are only contributions of the planar diagrams. Taking $S_f=0$ 
will then give only the contributions with at most one quark loop. Hence our 
proposal reproduces \cite{arcafehe1}, but in a large $N$ context. 

Another way to study this system using only the original DV proposal 
can be found by gauging the flavor symmetry. For this we consider a 
$\SU(N_c)\times \SU(N_f)$ super Yang-Mills with bifundamentals and 
adjoint Higgs fields for each of the groups. The effective superpotential 
for this model can be calculated by an ADE type 2-matrix model as in 
\cite{dv2,ade}. 
The form of the effective superpotential has roughly the form \eqref{weffprop} above. 
The $\SU(N_c)$ theory can be extracted by putting the $\SU(N_f)$ in a classical vacuum. 
Note that in this system $S_f=g_sM_f$ becomes the gluino condensate in the $\SU(N_f)$ factor. 
The $\SU(N_f)$ Higgs field then will play the role the mass matrix for the quarks. 
This can be accomplished by adding and putting the $\SU(N_f)$ glueball field 
$S_f$ to zero. As $S_f=g_sM_f$ in the matrix model this corresponds 
nicely to what we had above. This system can be understood in terms of 
string theory by wrapping D5-branes on two different 2-spheres intersecting at a point, 
where one of the two spheres is taken very large, freezing out the dynamics 
of the D-branes wrapping it. From this point of view this was also studied 
recently in \cite{ooku}. 

This paper is organized as follows. 
In \secref{sec:matrix} we discuss the matrix model approach and show in more 
detail the genus expansion for the different proposals. 
In \secref{sec:a2} we study the 2-matrix model of type $A_2$ and relate 
it to the flavor model. In \secref{sec:gaussian} we work out explicitly 
the model with quadratic superpotential and compare with another 
approach using meson potentials. In \secref{sec:concl} we conclude with 
same discussion.

\section{Flavors and Matrix Models}
\label{sec:matrix}

\subsection{Matrix Models With Flavors}

We consider an $\SU(N_c)$ gauge theory with $N_f$ pairs of quark fields 
$Q_i$ and $\tilde Q_i$, $i=1,\cdots,N_f$, in the fundamental and the anti-fundamental of 
$\SU(N_c)$ respectively. Note that the gauge superfield $W_\alpha$ contains a 
complex scalar $\Phi$ in the adjoint representation of the color group. 

The Lagrangian density of this theory is given by 
\begin{equation}
  L = \dint{d^4\theta}\Tr(\bar Q_i\e^V Q_i+\tilde Q_i\e^V\bar{\tilde Q}_i)
   + \dint{d^2\theta}\Tr(W(\Phi,Q,\tilde Q)+\tau W^\alpha W_\alpha),
\end{equation}
with superpotential given by 
\begin{equation}
  W(\Phi,Q,\tilde Q) = W(\Phi) + \Phi Q\tilde Q - Q m\tilde Q,
\end{equation}
where the first term is a polynomial tree level superpotential for the 
adjoint Higgs field $\Phi$ and $m$ is the mass matrix for the flavors. 
Classically the eigenvalues of the Higgs field and the quarks will be at 
critical points of this tree level superpotential. This breaks the gauge 
group to $\times_I\SU(N_I)$ where $N_I$ is the number of eigenvalues of 
$\Phi$ at the critical point $I$-th critical point. Quantum mechanically 
there will be a gluino condensate $S_I$ for each of the factors. The goal 
is to find the effective superpotential for these condensates. 

The effective superpotential can be calculated according to the proposal of 
Dijkgraaf and Vafa, using matrix models. Here we take the superpotential 
of the gauge theory reduced to a point in superspace, replacing the number 
of colors $N_c$ by the matrix model size $M$ of the number of flavors 
$N_f$ by $M_f$. So the fields are an $M\times M$ matrix $\Phi$, an 
$M\times M_f$ matrix $Q$ and a $M_f\times M$ matrix $\tilde Q$. 
The partition sum of the matrix model is given by 
\begin{equation}
  Z = \e^{\CF} = \frac{1}{Vol(G)}\dint{\CD\Phi\CD Q\CD\tilde Q}\e^{-\frac{1}{g_s}\Tr W(\Phi,Q,\tilde Q)},
\end{equation}
where $W$ is the same function as above, but now of the matrix variables, 
and $G$ is the unbroken gauge group. 

We denote by $g$ the genus of the diagram, and by $h$ the number of quark loops. 
By this we mean the number of boundary components coupling to the 
flavor matrices. Note that this is not the same as the number of holes 
(i.e. the number of holes coupling to the gauge multiplet). 
The dependence of the free energy of the parameters $g_s$, $M$, and 
$M_f$ can easily be extracted from the topology of the diagrams and 
can be written as an expansion in the genus and the number of 
quark loops as
\begin{equation}
  \CF = \sum_{g,h}g_s^{2g+h-2}M_f^h\CF_{g,h}(g_sM)
  = \sum_{g,h}g_s^{2g-2}(g_s M_f)^h\CF_{g,h}(g_sM). 
\end{equation}
Let us therefore introduce parameters $S\equiv g_sM$ and $S_f\equiv g_sM_f$. 

We propose, analogous to DV, that the effective superpotential can be 
found from an expression of the form
\begin{equation}\label{weff}
  W_{eff}(S) = N_c\frac{\partial\CF_0}{\partial S}(S,S_f=0)+N_f\frac{\partial\CF_0}{\partial S_f}(S,S_f=0),
\end{equation}
where $\CF_0$ is the planar contribution to the free energy for genus 0. 
This is a generalization of the pure gauge theory studied by Dijkgraaf and Vafa. 

The planar contributions can be found from a large $M$ expansion of the 
matrix model. All planar diagrams are summed by taking both the rank 
of the gauge group and the number of flavors to infinity, 
\begin{equation}
  M,\,M_f\to\infty,\qquad g_s\to 0,\qquad S= g_sM,~S_f= g_sM_f\quad\mbox{finite}.
\end{equation}
From the expansion above we see that this picks out the genus zero contribution. 

Let us relate the proposal above to the calculation of the matrix model 
in a large $M$ limit taking $M_f=N_f$ fixed, as in \cite{arcafehe1} 
and many papers following it. 
From the above expansion we find in the large $M$ limit 
\begin{equation}
  N_c\frac{\partial\CF_0}{\partial S}+N_f\frac{\partial\CF_0}{\partial S_f} 
  = \sum_{h\geq0}\biggl(N_cS_f^h\frac{\partial\CF_{0,h}(S)}{\partial S}+hN_fS_f^{h-1}\CF_{0,h}(S)\biggr). 
\end{equation}
Taking $S_f=0$ there are only two surviving terms: 
in the first sum we only get the $\CF_{0,0}$ contribution, while in 
the second sum we only retain $\CF_{0,1}$.
Therefore the effective superpotential is given by 
\begin{equation}
  W_{eff}(S) = N_c\frac{\partial\CF_{0,0}}{\partial S}(S)+ N_f \CF_{0,1}(S) 
  = N_c\frac{\partial\CF_{\chi=2}}{\partial S}(S) + \CF_{\chi=1}(S). 
\end{equation}
We find the same expression as \cite{arcafehe1}. The advantage of our proposal 
is that in keeping $M_f=N_f$ finite, the contribution at $\chi=1$ is 
subleading in $g_s$ or equivalently $1/M$. Therefore one cannot completely 
rely on large $M$ expansion in the matrix model, which makes it harder 
to do actual computations, especially if one wants to consider general 
tree level superpotentials for the adjoint Higgs field. 

We should note that, as opposed to the proposal of \cite{mcgre}, we do not 
relate $M_f$ to $N_f$. Particularly, the ration $M_f/M$ is not fixed in the 
large $M$ limit. This is necessary, as we want to take $S_f=0$ at the end. 
Indeed, the matrix model quantities $M$ and $M_f$ are completely unrelated 
to the gauge theory quantities $N$ and $N_f$.

\subsection{Loop Equations}

Let us first consider the matrix model with a large number of flavors, as described above. 
To calculate the matrix model integral, we integrate out the fundamentals. 
We are then left with an integral over $\Phi$, 
\begin{equation}
  Z = \dint{\CD\Phi}\e^{-\frac{1}{g_s}\Tr V(\Phi)},
\end{equation}
where 
\begin{equation}
  V(\Phi) = W(\Phi) + S_f\log(\Phi-m).
\end{equation}
We take the large $M,M_f$ limit with 
$g_s\to0$ such that the 't Hooft coupling $S=g_s M$ and $S_f=g_sM_f$ is fixed. 

As is usual we introduce the resolvent defined as 
\begin{equation}
  \omega(x) = \frac{1}{M}\Tr\frac{1}{x-\Phi}. 
\end{equation}
We can then derive the loop equation 
\begin{equation}
  \doint[_C]{\frac{dz}{2\pi i}}\frac{V'(z)}{x-z}\vev{\omega(z)} = S\vev{\omega(x)^2},
\end{equation}
as a Schwinger-Dyson equation for the matrix model. Here we have to be 
careful that $C$ is a cycle enclosing all the cuts but not the point $x$ or $m$. 
The latter is because $V'(z)$ has a pole at $z=m$. 
This loop equation can also be written in the form 
\begin{equation}
  S\vev{\omega(x)^2} - V'(x)\vev{\omega(x)}+\frac{1}{4S}f(x) = 0,
\end{equation}
where 
\begin{equation}
  f(x) = \frac{4S}{M}\biggl\langle\Tr\frac{V'(x)-V'(\Phi)}{x-\Phi}\biggr\rangle.
\end{equation}
Note that if $V$ were polynomial, this would be a polynomial in $x$. 
Also, in the large $M$ limit we have 
$\vev{\omega(x)^2}= \vev{\omega(x)}^2+\CO(\frac{1}{M^2})$. Therefore 
we can replace $\omega(x)$ by its expectation value. Introducing 
 $y(x)=-2S \omega(x)+V'(x)$ we find the spectral curve 
\begin{equation}
  y^2=V'(x)^2-f(x). 
\end{equation}
Unlike the usual matrix model however, the functions $V'(x)$ and $f(x)$ have a pole 
at $x=m$. The pole term in $f(x)$ is given by 
\begin{equation}
  \frac{4S}{M}\biggl\langle\Tr\Biggl(\frac{1}{x-\Phi}\biggl(\frac{S_f}{x-m}-\frac{S_f}{\Phi-m}\biggr)\Biggr)\biggr\rangle
  = \frac{4SS_f}{x-m}\omega(m).
\end{equation}

As a result a multi-cut solution has the form 
\begin{equation}
  y(x) = \biggl(M(x)+\frac{\gamma}{x-m}\biggr)\sqrt{\prod_i(x-x_i)},
\end{equation}
where the cuts lie along the intervals $[x_{2i-1},x_{2i}]$, $M(x)$ is a polynomial  
with degree determined by the number of cuts and the degree of $W$, and 
$\gamma$ is determined by the pole at $x=m$ to be 
\begin{equation}
  \gamma=\frac{S_f}{\sqrt{\prod_i(m-x_i)}}. 
\end{equation}
Note that this coefficient is also such that the resolvent 
\begin{equation}
  2S\omega(x) = W'(x)+\frac{S_f}{x-m}-y(x),
\end{equation}
has no pole at $x=m$. 

Rather than pursuing this direction, we want to proceed to a 
different method, relating the model to a special limit of a 2-matrix model. 
As it turns out several aspects of the solution can be seen much more 
clearly in this approach.

\section{Flavors From the 2-Matrix Model}
\label{sec:a2}

In this section we discuss the use of the 2-matrix model to calculate the 
effective superpotential. We will interpret the two gauge group 
as the color and flavor symmetry group, which now are both gauged. 
To find the $\SU(N_c)$ theory with fundamental flavors we 
take a limit forcing the flavor sector into a classical vacuum.

\subsection{The $A_2$ Model}

We propose a description of the flavor model in terms of certain multi-matrix models. 
In particular, the $A_2$ model. The idea is as follows. The $A_2$ model 
is based on a quiver with 2 nodes, both corresponding to a gauge group $\U(N_i)$. 
We have two bifundamentals, which we call $Q=Q_{12}$ and $\tilde Q=Q_{21}$. The 
tree level superpotential is given by 
\begin{equation}
  \Tr_1 W_1(\Phi_1)+\Tr_2 W_2(\Phi_2)+\Tr_1 \tilde Q\Phi_1 Q-\Tr_2 Q\Phi_2\tilde Q. 
\end{equation}

Let us discuss the solution of this model. Part of the following discussion, 
and some more details, can also be found in \cite{dv2}. 
We can write the matrix integrals in terms of eigenvalues $\lambda_{i,I}$ 
of the matrices $\Phi_i$. 
We introduce the resolvents for the two adjoints, 
\begin{equation}
  \omega_i(x) = \frac{1}{M_i}\Tr_i\frac{1}{x-\Phi_i} = \frac{1}{M_i}\sum_{I}\frac{1}{x-\lambda_{i,I}}.
\end{equation}
After integrating out the bifundamentals $Q,\tilde Q$, we obtain \cite{dv2}
\begin{equation}
  \dint{\prod_{i,I}d\lambda_{i,I}}\prod_{(i,I)\neq (j,J)}(\lambda_{i,I}-\lambda_{j,J})^{C_{ij}}
  \e^{-\frac{1}{g_s}\sum_{i,I}W_i'(\lambda_{i,I})}.
\end{equation}
with $C_{ij}=2\delta_{i,j}-\delta_{i,j-1}-\delta_{i,j+1}$ the Cartan matrix of the $A_2$ model. 
The forces, or equivalently equations of motion, can be written 
\begin{equation}
  y_i(x) = -g_s\partial_{\lambda_{i,I}} S(x) = W_i'(x)-C_{ij}S_j\omega_i(x),
\end{equation}
where $S_i=g_sM_i$ are the 't Hooft couplings. 
It is convenient to introduce two scalar fields $\varphi_i(x)$ on the $x$-plane 
related to the forces as $\partial\varphi_i(x) = y_i(x)dx$. The matrix integral 
can in fact be related to the rational CFT of type $A_2$ in terms of these scalar fields. 
This CFT is known to have a structure of a $W_3$ algebra formed by 2 currents 
$\CW^{(s)}(x)$ of spins $s=2,3$. These currents also exist in the matrix model \cite{ade}. 

To write expressions for the currents, it is convenient to use a slightly 
different basis for the forces \cite{dv2}, 
\begin{eqnarray}
  a_1(x) &=& t_1(x)-S_1\omega_1(x),\nonumber\\
  a_2(x) &=& t_2(x)+S_1\omega_1(x)-S_2\omega_2(x),\\
  a_3(x) &=& t_3(x)+S_2\omega_2(x), \nonumber
\end{eqnarray}
where the $t_i$ are the classical contributions 
\begin{eqnarray}
  t_1(x) &=& \frac{2}{3}W_1'(x)+\frac{1}{3}W_2'(x),\nonumber\\
  t_2(x) &=& -\frac{1}{3}W_1'(x)+\frac{1}{3}W_2'(x),\\
  t_3(x) &=& -\frac{1}{3}W_1'(x)-\frac{2}{3}W_2'(x). \nonumber
\end{eqnarray}
These are not independent, but satisfy $\sum_ia_i=0=\sum_it_i$. 
The two spin $s$ currents can then be given by \cite{ade}
\begin{equation}
  \CW^{(s)}(x) = \frac{(-1)^s}{s}\sum_{i=1}^3a_i(x)^s,\qquad s=2,3.
\end{equation}
They have a classical contribution $\CW^{(s)}_{cl}(x)$ which can be found by 
replacing the $a_i$ with the $t_i$. 
These currents satisfy loop equations of the form 
\begin{equation}
  \doint[_C]{\frac{dz}{2\pi i}}\frac{1}{x-z}\vev{\CW^{(s)}(z)} = 0.
\end{equation}
where $C$ is a cycle enclosing all the eigenvalues but not the point $x$. 
The loop equations imply that 
\begin{equation}
  \vev{\CW^{(s)}(x)}-\CW^{(s)}_{cl}(x) = w^{(s)}(x),
\end{equation}
where $w^{(s)}(x)$ are regular functions of $x$. 
If the $W_i(x)$ are polynomials of degree $n+1$, one can show that 
$w^{(s)}$ is a polynomial of degree $ns-n-1$. In the large $M$ limit 
the expectation values factorize, so we do not have to distinguish between the 
resolvent as an operator and its expectation value in the expression for the current. 

The quantum spectral curve of this model can be written 
\begin{equation}
  \prod_{i=1}^3(y-a_i(x))=0. 
\end{equation}
Using the currents we can write this as a deformation of the classical 
curve --- the above curve with the $a_i$ replaced by the $t_i$ --- 
\begin{equation}
  \prod_{i=1}^3(y-t_i(x))+yw^{(2)}(x)+w^{(3)}(x) = 0.
\end{equation}
This quantum curve can now be conveniently used to study the large $M$ behavior 
of the model. 

\begin{figure}[ht]
\begin{center}
\begin{picture}(300,210)
\put(0,0){
}
\put(130,195){$P$}
\put(130,115){$Q$}
\put(130,34){$R$}
\put(260,200){sheet 1}
\put(260,120){sheet 2}
\put(260,40){sheet 3}
\put(110,174){$C_{1,I}$}
\put(140,174){$C_{12,I}$}
\put(150,94){$C_{2,I}$}
\put(45,170){$A_{1,I}$}
\put(205,170){$A_{12,I}$}
\put(235,90){$A_{2,I}$}
\end{picture}
\end{center}
\caption{The three sheets of the $A_2$ curve, and the three cuts connecting them. 
The points $P,Q,R$ are the points at regularized infinity.}
\label{fig:sheets}
\end{figure}
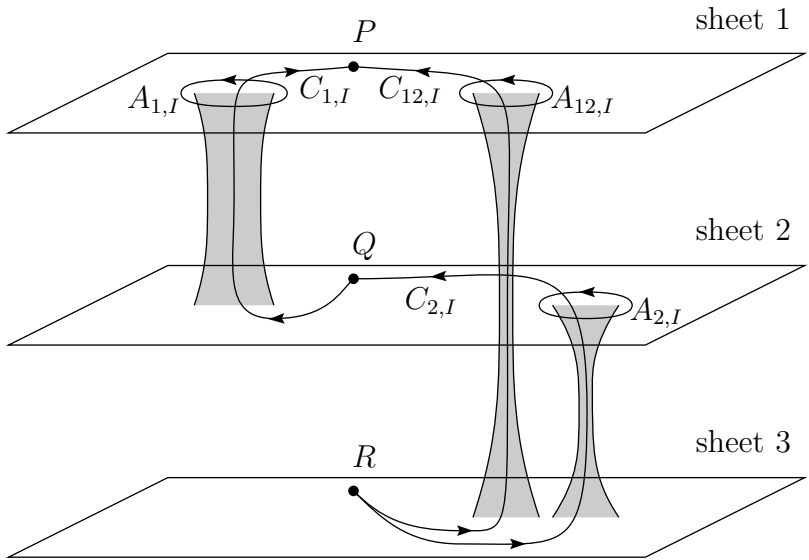

\subsection{Free Energy at Large $M$ and the Effective Superpotential}

We will now study the large $M$ limit by looking at the shape of the quantum curve. 
Let us first study the classical equations of motion of the system. 
In the following we will assume that there are no simultaneous 
solutions to $W_1'(\lambda)=0$ and $W_2'(\lambda)=0$. This is not a 
strong condition and will be sufficient for the rest of this paper. 

We can use the $\U(N_1)\times\U(N_2)$ symmetry to diagonalize the 
matrices $\Phi_i$; their eigenvalues will be denoted $\lambda_{i,I}$. 
For the corresponding blocks $Q_{IJ}$ and $\tilde Q_{JI}$ in $Q$ and 
$\tilde Q$ the equations of motion give 
\begin{equation}
  (\lambda_{1,I}-\lambda_{2,J})Q_{IJ} = 0,\qquad
  (\lambda_{1,I}-\lambda_{2,J})\tilde Q_{JI} = 0.
\end{equation}
This implies that the blocks $Q_{IJ}$ and $\tilde Q_{JI}$ 
are nonzero only if $\lambda_{1,I}=\lambda_{2,J}$. We can therefore safely 
reduce the discussion to individual eigenvalues and henceforth suppress 
the extra index $I$. 

The remaining $\Phi_i$ equations of motion are given by 
\begin{equation}\label{eom}
  W_1'(\lambda_1) = -Q\tilde Q,\qquad
  W_2'(\lambda_2) = \tilde QQ,
\end{equation}
where the left-hand sides have to be interpreted as diagonal matrices. 
There are three types of solutions to these equations. 

When a block in $Q\tilde Q$ is zero, we find a solution for a critical 
point of the first node, $W_1'(\lambda_1)=0$. 
We will call these critical points $e_{1,I}$. Similarly, we have solutions 
$e_{2,I}$ corresponding to the critical points of the second node, 
$W_2'(\lambda_2)=0$. 
It remains to study the critical points when the eigenvalues are equal, 
$\lambda_1=\lambda_2\equiv\lambda_{12}$. It follows from the equations of 
motion \eqref{eom} and some elementary algebra that 
$Q\tilde Q=\tilde QQ$.\footnote{Assume, w.l.o.g., that $W_1'(\lambda_{12})\neq0$. 
Then we can write, after a basis transformation, 
$Q=\pmatrix{Q_1&0}$ and $\tilde Q=\pmatrix{\tilde Q_1\cr\tilde Q_2}$, 
with $Q_1$ and $\tilde Q_1$ invertible matrices. We find $\tilde QQ\neq0$ 
and, using \eqref{eom}, it must be that $Q=Q_1$. Therefore 
$\tilde QQ=Q\inv (Q\tilde Q)Q=Q\tilde Q$, 
as $Q\tilde Q$ is central.} 
Therefore, these eigenvalues satisfy 
\begin{equation}
  W_1'(\lambda_{12})+W_2'(\lambda_{12}) = 0.
\end{equation}
The solutions to this equation will be called $e_{12,I}$. 

The three different types of classical solutions can be related to the intersections 
of the three sheets comprising the classical curve. The intersection of the 
$i$-th and $j$-th sheet is given by the solutions to $t_i(x)-t_j(x)=0$. 

Quantum mechanically, the eigenvalues for $\Phi_i$ will spread around these 
classical solutions. In the large $M$ limit they will form cuts in the complex $x$-plane. 
The contours around the cuts will be denoted $A_{i,I}$ and $A_{12,I}$. 
The resulting quantum curve is sketched in \figref{fig:sheets}; the 3 sheets 
correspond to the solutions $y=a_i(x)$. We also indicated the positions 
of the cuts around the different classical solutions. 

The number of eigenvalues, or rather the 't Hooft couplings $S_{i,I}=g_sM_{i,I}$, 
can be measured by the contour integrals around the cuts, 
\begin{equation}
  S_{i,I} = -\frac{1}{4\pi i}\oint_{A_{i,I}}y(x)dx,\quad i=1,2;\qquad
  S_{12,I} = -\frac{1}{4\pi i}\oint_{A_{12,I}}y(x)dx.
\end{equation}
Integrating at infinity on the first and third sheet, we find 
that the total 't Hooft couplings $S_i=g_sM_i$ are given by the sums 
\begin{equation}\label{sdecomp}
  S_1 = \sum_IS_{1,I}+\sum_IS_{12,I}, \qquad
  S_2 = \sum_IS_{2,I}+\sum_IS_{12,I}. 
\end{equation}

For our purpose we need to solve the 2-matrix model in the large $M$ limit. 
At large $M$ we can replace the sums over eigenvalues by the integral over 
eigenvalue distributions $\rho_1(\lambda_1)$ and $\rho_2(\lambda_2)$ respectively. 
They can be found from the discontinuity in the corresponding resolvents. 
The large $M$ limit of the free energy can be found from the saddle point approximation, 
\begin{equation}
\CF_0(S_1,S_2) = - S_i\dint{d\lambda_i}\rho_i(\lambda_i)W_i(\lambda_i) 
  + \frac{1}{2}C_{ij}S_iS_j\dint{d\lambda_id\lambda_j}\rho_i(\lambda_i)\rho_j(\lambda_j) \log(\lambda_i-\lambda_j). 
\end{equation}

As in \cite{dv1}, the free energy can be found from the curve by considering the 
change of the density when we remove an eigenvalue from the cuts. 
When we remove an eigenvalue from the cut at $x=e_{i,I}$, the density changes by  
\begin{equation}
  \delta_{i,I}(S_{i}\rho_i)(\lambda_{i}) = \delta S_{i,I}\delta(\lambda_{i}-e_{i,I}), \qquad
  i=1,2.
\end{equation}
Using this we find, for $i=1,2$, a variation 
\begin{equation}
  \frac{\partial\CF_0}{\partial S_{i,I}} = 
  -W_i(e_{i,I}) + C_{ij}S_j\dint{d\lambda_j}\rho_j(\lambda_j)\log(\lambda_j-e_{i,I})
  = \int_{e_{i,I}}^\infty (a_i(x)-a_{i+1}(x))dx.
\end{equation}
As $y(x)=a_i(x)$ on the $i$-th sheet, we can write this as a contour integral over  
the noncompact curve $C_{i,I}$ going through the appropriate cut, 
as indicated in \figref{fig:sheets}, 
\begin{equation}
  \frac{\partial\CF_0}{\partial S_{i,I}} = \int_{C_{i,I}}y(x)dx.
\end{equation}
As the curves are noncompact we should regularize these integrals 
by ending the curves at finite points $P$, $Q$ and $R$ on the three sheets 
respectively. A similar discussion holds when we remove an eigenvalue 
at $e_{12,I}$. 
We now find the integral over $a_1(x)-a_3(x)$, which can be  
written as an integral over the noncompact curve $C_{12,I}$ 
connecting the first and third sheet, 
\begin{equation}
  \frac{\partial\CF_0}{\partial S_{12,I}} = \int_{C_{12,I}}y(x)dx. 
\end{equation}

Let us now relate these to the effective superpotential of the quiver gauge theory. 
The original $\SU(N_1)\times\SU(N_2)$ gauge group will be broken by the distribution 
of the eigenvalues of the adjoint Higgs fields around the different classical vacua. 
The ranks therefore decompose analogously to 
\begin{equation}\label{ndecomp}
  N_1 = \sum_IN_{1,I}+\sum_IN_{12,I},\qquad
  N_2 = \sum_IN_{2,I}+\sum_IN_{12,I},
\end{equation}
analogous to \eqref{sdecomp}. 
Here the $\SU(N_{12,I})$ factors are embedded diagonally in both original group factors. 
The effective potential for the $A_2$ quiver theory should now be given 
in terms of the matrix theory quantities by 
\begin{equation}\label{weffa2}
  W_{eff} = N_{1,I}\frac{\partial\CF_0}{\partial S_{1,I}}
  + N_{2,I}\frac{\partial\CF_0}{\partial S_{2,I}}
  + N_{12,I}\frac{\partial\CF_0}{\partial S_{12,I}}. 
\end{equation}

\subsection{Decoupling the Flavor Group}

To relate the $A_2$ model to the flavor model, we identify the
group $\SU(N_2)$ with the flavor group. To decouple this group we 
need to take this sector in a classical vacuum. Doing so, the 
adjoint $\Phi_2$ of this node will play the role of the mass matrix $m$. 
To get the right masses we take the tree level superpotential 
of the flavor node such that 
\begin{equation}
  W_2'(\Phi_2) = \kappa\prod_{a=1}^{N_f}(\Phi_2-m_a).
\end{equation}
We will then take  $\kappa\to\infty$. This ensures that the flavor 
node will not add any quantum corrections, so that indeed $\Phi_2$ will be 
constrained to the critical points $m_a$. 

So we have to consider the limit $\kappa\to\infty$ of the cubic Riemann surface. 
In this limit the Riemann surface degenerates, and the third sheet 
will decouple. Note that for large $\kappa$, all the cuts in sheet 3 
corresponding to $y=a_3(x)$ are around the points $x=m_a$, as $W_2'$ dominates. 
For large $\kappa$ around $x=m_a$, the resolvent $S_2\omega_2(x)$ will have the form 
\begin{equation}
  S_2\omega_2(x)\sim \frac{1}{2}\kappa_a(x-m_a)-\frac{1}{2}\sqrt{\kappa_a^2(x-m_a)^2-4\kappa_aS_{2,a}},
\end{equation}
where $\kappa_a$ is proportional to $\kappa$. $\kappa_a$ and $S_{2,a}$ in the above equation 
will depend on $x$, but have finite values at $x=m_a$. Taking large $\kappa$ we have 
\begin{equation}
  S_2\omega_2(x)\sim \frac{1}{2}\kappa_a(x-m_a)-\frac{1}{2}\sqrt{\kappa_a^2(x-m_a)^2-4\kappa_aS_{2,a}} 
  = \frac{S_{2,a}}{x-m_a}+\CO\biggl(\frac{1}{\kappa}\biggr). 
\end{equation}
Therefore in the limit $\kappa\to\infty$ we have 
\begin{equation}
  S_2\omega_2(x) = \sum_{a=1}^{N_f} \frac{S_{2,a}}{x-m_a}. 
\end{equation}
The above arguments would also allow extra regular terms, but the asymptotics 
$\omega_2(x)\sim\frac{1}{x}$ for large $x$ does not allow them. Therefore 
the above form is exact in the limit $\kappa\to\infty$. Of course this was expected; 
the eigenvalues of $\Phi_2$ are all in their classical vacua at $\lambda_2=m_a$. 

We see that in the limit the branch cut dissapears, therefore 
the equation for the Riemann surface factorizes in this limit. 
We should then be interested in the part coming from the other two sheets. 
We have to change the coordinate $y$ to absorb the large $\kappa$ part. 
The proper shifted coordinate on the Riemann surface can be given by 
\begin{equation}
  \tilde y = 2y+a_3(x). 
\end{equation}
We shift the $a_{1,2}$ by the same amount, 
\begin{equation}
  \tilde a_1 = 2a_1+a_3,\qquad \tilde a_2 = 2a_2+a_3. 
\end{equation}
Note that we now have $\tilde a_1+\tilde a_2=0$. The equation 
$(y-a_1)(y-a_2)=0$ for the decoupled first two sheets becomes 
\begin{equation}
  (\tilde y-\tilde a_1)(\tilde y-\tilde a_2)
  = \tilde y^2-(W_1'(x)+S_2\omega_2(x))^2+f(x). 
\end{equation}
The quantum correction 
\begin{equation}
  f(x) = -4S_1^2\omega_1^2+4S_1\omega_1W_1'+4S_1S_2\omega_1\omega_2,
\end{equation}
is a rational function with single poles at $x=m_a$. This can be seen by 
writing this in terms of the quantum contribution to the spin 2 current, 
\begin{equation}
   S_1^2\omega_1(x)^2-S_1S_2\omega_1(x)\omega_2(x)-S_1\omega_1(x)W_1'(x) 
   = w^{(2)}(x) - S_2^2\omega_2(x)^2+S_2\omega_2(x)W_2'(x), 
\end{equation}
where $w^{(2)}(x)$ is a polynomial function of $x$. 
We then have 
\begin{equation}
  \oint_C\frac{dz}{2\pi i}\frac{1}{x-z}f(z) = 0, 
\end{equation}
where now $C$ is a cycle containing all the eigenvalues of $\Phi_1$, but not the 
eigenvalues $m_a$ and the point $x$. We then find that $f(x)$ 
is regular away from $x=m_a$. As it only has at most a single power 
of $\omega_2$, it will have single poles at $x=m_a$. Near 
$x=m_a$ it can be approximated 
\begin{equation}
  f(x) \approx 4S_1S_{2,a}\omega_1(m_a)\frac{1}{x-m_a}. 
\end{equation}
In fact, this decoupled curve is precisely the one we found in the last section 
governing the matrix model with a large number of flavors. This shows that 
indeed as expected the two methods are equivalent. 

Let us change some notation appropriate for the present setting. We will 
denote $S_I=S_{1,I}$, $S_{f,a}^+=S_{2,I}$ and $S_{f,a}^-=S_{12,I}$, and 
similar notations for the $N_I$. 
Let us now see what form the derivatives of the free energy reduce to. 
First we note that the $a_1-a_2=\frac{1}{2}(\tilde a_1-\tilde a_2)$, which is 
half the difference of $\tilde y$ on the two sheets. Therefore we can write the 
variation of the tree level superpotential with respect to the moduli 
$S_{1,I}$ as 
\begin{equation}
  \frac{\partial\CF_0}{\partial S_{I}} = \frac{1}{2}\int_{C_{I}} \tilde y(x)dx, 
\end{equation}
where $C_I$ is the noncompact curve in the decoupled double cover 
corresponding to $C_{1,I}$, running from $Q$ to $P$. 

For the variation with respect to $S_{f,a}^+$, we note that the 
$a_2-a_3=\frac{1}{2}(\tilde a_2-3a_3)$. We have $3a_3 =W_1'+2W_2'+3S_2\omega_2$, 
but $W_2'$ is independent of $S$ while the last term vanishes at $S_2=0$. 
Therefore we can write 
\begin{equation}
  \frac{\partial\CF_0}{\partial S_{f,a}^+} = \int_{m_a}^{\infty} (a_2(x)-a_3(x))dx 
  = \frac{1}{2}\int_{m_a}^{Q} (\tilde y(x)-W_1'(x))dx + constant.  
\end{equation}
We can safely ignore the extra constant for the calculation of the superpotential. 
In fact it is the contribution of the part of $C_{2,I}$ lying on the 
decoupled third sheet, and represents the residue of the second node. 
Similarly, for the degenerate cuts connecting the first and the second 
sheet we find at $S_2=0$ 
\begin{equation}
  \frac{\partial\CF_0}{\partial S_{f,a}^-} = \int_{m_a}^{\infty} (a_1(x)-a_3(x))dx 
  = \frac{1}{2}\int_{m_a}^{P} (\tilde y(x)-W_1'(x))dx + constant.  
\end{equation}
The constants are related to the dynamics of the second node. As we 
are not interested in these, we can simply substract them. As they 
do not depend on the glueball expectation values this can be safely done. 

For the flavor model we are interested in the limit $S_f=0$. The only 
relevant parameters are $S_I\equiv S_{1,I}$. In the expressions 
above for the derivatives of the free energy 
we should substitute the solution for $\tilde y$ at $S_2=0$. 
This is simply the solution of the 1-matrix model with tree level 
superpotential $W_1(x)$, and moduli determined by $S_{I}$. 

The effective superpotential can now be calculated from \eqref{weffa2}
with the proper interpretation of the factors. Note that 
\eqref{ndecomp} gives the decomposition of the ranks as 
\begin{equation}\label{nfdecomp}
  N_c = \sum_I N_I+\sum_a N_{f,a}^-,\qquad
  N_f = \sum_{a,\pm} N_{f,a}^\pm.
\end{equation}
Hence we find that the different choices for $\epsilon_a=\pm1$ 
give rise to different unbroken flavor symmetries.

\subsection{Remarks on Seiberg Duality}

The $A_2$ curve has a natural action of the $A_2$ Weyl group $S_3$. 
The part of this group compatible with our limit $\kappa\to \infty$ 
decoupling the third sheet is a $\Z_2$ symmetry interchanging 
sheets one and two. 
Looking at the action of this $\Z_2$ on the cuts 
they act on the 't Hooft couplings as 
$\tilde S_{I}=-S_{1,I}$, $\tilde S_{2,I}=S_{12,I}$ and $\tilde S_{12,I}=S_{2,I}$. 
The curve after the $\Z_2$ action describes a similar gauge system but 
with different parameters. 
In terms of these dual variables the effective superpotential becomes 
\begin{equation}
  N_I\frac{\partial\CF_0}{\partial S_I}+N_{f,a}^+\frac{\partial\CF_0}{\partial S_{f,a}^+}+N_{f,a}^-\frac{\partial\CF_0}{\partial S_{f,a}^-} 
  = -N_I\frac{\partial\CF_0}{\partial \tilde S_I'}+N_{f,a}^-\frac{\partial\CF_0}{\partial \tilde S_{f,a}^+}+N_{f,a}^+\frac{\partial\CF_0}{\partial \tilde S_{f,a}^-}.
\end{equation}
We read off from this that this can be interpreted as the effective superpotential 
for a dual theory with $\tilde N_I=-N_I$ while $\tilde N_{f,a}^\pm=N_{f,a}^\mp$. 
This implies 
\begin{equation}
  \tilde N_c = \sum_I \tilde N_I+\sum_a \tilde N_{f,a}^- = -\sum_I N_I+\sum_a N_{f,a}^+ = N_f-N_c,\qquad
  \tilde N_f = N_f. 
\end{equation}
Note that for this to have a gauge theory interpretation we need 
$N_f>N_c$, as the number of colors is always positive. This is exactly the 
transformation of Seiberg's electric-magnetic duality \cite{sei}. 
Indeed this Weyl group symmetry has been connected to 
Seiberg-like duality in the context of D-branes on Calabi-Yau geometries 
with ADE geometry in \cite{cafiinkava}. 

Note that the quantum curve has two types of degenerate cuts connecting the 
third ``flavor'' sheet, corresponding to $\epsilon_a=\pm1$. 
These have a different physical interpretation. This can be seen form the 
behavior of the $N_c$ dependence. The duality described above 
interchanges these two cuts. In the Seiberg duality, the magnetic model 
also is slightly different, as it has extra gauge invariant fields, 
and an extra term in the tree level superpotential. Furthermore the original 
Seiberg duality has massless quarks, so we need to take a limit $m\to0$.

\section{The Gaussian Model}
\label{sec:gaussian}

\subsection{Matrix Model Calculation}

Let us for specifity consider the simplest example of $\U(N_c)$ super Yang-Mills with 
$N_f$ massive flavors and an adjoint, with a gaussian tree level superpotential for the adjoint. 
That is, we take 
\begin{equation}
  W_1(\Phi_1) = \frac{g}{2}\Phi_1^2.
\end{equation}
The superpotential $W_2$ of the second node will be taken in terms of the masses $m_a$ 
for the flavors as above. 

The tree level superpotential has only a single vacuum at $\Phi_1=0$. Hence there 
will be only a single glueball $S_{1,I}=S$. For $S_2=0$ the solution for $y(x)$ is 
has the well known half circle form, 
\begin{equation}
  y(x) = \sqrt{W_1'(x)^2-f(x)} = \sqrt{g^2x^2-4gS}. 
\end{equation}
(We leave out the tilde as we are only considering the decoupled system).

We take the noncompact cycle $C=C_I$, connecting the point running from $Q$ to $P$ 
passing through the $S$ cut. We will take a regularization where $P$ and $Q$ are taken at 
$\tilde\Lambda$. We then have, at $S_2=0$, $S_1=S$, 
\begin{equation}
  \frac{\partial\CF_0}{\partial S} = \frac{1}{2}\int_C y(x)dx = -S\log \frac{S}{g\tilde\Lambda^2}+S.
\end{equation}
For the derivative with respect to $S_{f,a}^+$ at $S_f=0$ we have to take 
the integral from the point $m_a$ to either $P$ or $Q$. We find, up to a 
$S$ independent constant, 
\begin{eqnarray}
  \frac{\partial\CF_0}{\partial S_{f,a}^+} &=& \frac{1}{2}\int_{m_a}^{Q} (y(x)-W_1'(x))dx \\
 &=&  S\log\biggl(\frac{m_a}{\tilde\Lambda}\biggr)-\frac{1}{2}S + S\log\biggl(\frac{1}{2}+\frac{1}{2}\sqrt{1-4\alpha_aS}\biggr)-\frac{1}{4\alpha_a}\sqrt{1-4\alpha_aS}
  +\frac{1}{4\alpha_a},\nonumber\\
 &=&  S\log\biggl(\frac{m_a}{\tilde\Lambda}\biggr) - \frac{1}{2}S + S\log\gamma_{+}(\alpha_aS) + \frac{S}{2\gamma_+(\alpha_aS)},\nonumber
\end{eqnarray}
where $\alpha_a=\frac{1}{gm_a^2}$ and we have introduced the functions 
\begin{equation}
  \gamma_{\pm}(x) = \frac{1}{2}\pm\frac{1}{2}\sqrt{1-4x}. 
\end{equation}
Similarly, we can calculate the derivative with respect to $S_{f,a}^-$, 
for which we find
\begin{eqnarray}
  \frac{\partial\CF_0}{\partial S_{f,a}^-} &=& \frac{1}{2}\int_{m_a}^{P} (y(x)-W_1'(x))dx \\
 &=&  -S\log\biggl(\frac{S}{gm_a\tilde\Lambda}\biggr) + \frac{1}{2}S + S\log\gamma_{-}(\alpha_aS) + \frac{S}{2\gamma_-(\alpha_aS)},\nonumber
\end{eqnarray}
From our proposal \eqref{weff}, 
we can give the following expression for the effective superpotential, 
noting that $N_{1}=N_c-N_{f}^-$ due to \eqref{ndecomp}, 
\begin{equation}
  W_{eff}(S) = -N_cS\biggl(\log \frac{S}{\Lambda^{3}}-1\biggr)
  + \sum_{a=1}^{N_f} S\biggl(\log \gamma_{\epsilon_a}(\alpha_aS)
  +\frac{1}{2\gamma_{\epsilon_a}(\alpha_aS)}-\frac{1}{2}\biggr). 
\end{equation}
where $\epsilon_a=\pm1$ according to whether the integral was taken to 
$Q$ or $P$, and the scale $\Lambda$ is given in terms of the cutoff as 
$\Lambda^{3N_c}=\tilde\Lambda^{2N_c-N_f}g^{N_c}\det m$. 
This is precisely the form found by 
an explicit summation of the planar diagrams with up to one boundary 
\cite{arcafehe1,ohta}. 

The actual effective superpotential can be found by minimizing the above with respect 
to $S_1$, $\partial_SW_{eff}=0$. 
The critical points are determined by the solutions to the equation 
\begin{equation}\label{scrit}
  \biggl(\frac{S}{\Lambda^3}\biggr)^{N_c} = \prod_{a=1}^{N_f}\gamma_{\epsilon_a}(\alpha_aS).
\end{equation}
At a critical point the logarithmic terms in the effective superpotential 
cancel, and it can be written 
\begin{equation}\label{wcrits}
  W_{eff,crit} = \frac{2N_c-N_f}{2}S+\sum_{a=1}^{N_f}\frac{S}{2\gamma_{\epsilon_a}(\alpha_aS)},
\end{equation}
where $S$ is a solution to the equation above. 

For simplicity we will now consider the case where all masses are equal to $m$ 
and all $\epsilon_a=1$. This gives an equation for $S$ which can be written 
\begin{equation}
 \biggl( \frac{S}{\Lambda^3} \biggr)^{\frac{2N_c-N_f}{N_f}}-\biggl( \frac{S}{\Lambda^3} \biggr)^{\frac{N_c-N_f}{N_f}}+\frac{\Lambda^3}{m^2} = 0.
\end{equation}
Here we used that $S=0$ can not be a solution. The expression for the 
effective superpotential at the critical point is given by 
\begin{equation}
  W_{eff,crit} = \frac{\Lambda^3}{2}\Biggl((2N_c-N_f)\frac{S}{\Lambda^3} 
  + N_f\biggl(\frac{\Lambda^3}{S}\biggr)^{\frac{N_c-N_f}{N_f}}\Biggr),
\end{equation}
where for $S$ we should insert a solution to the above equation.

\subsection{Calculation Via Meson Potential}

Another way to calculate the answer is to integrate out $\Phi$ and express 
everything in terms of the meson field $X=\tilde QQ$. This procedure was 
used in \cite{arcafehe1} for the case of a single flavor, and in 
\cite{ohta} for more flavors. It is useful though for us to review it 
as it will give an interpretation of the two flavor cuts in terms of meson 
expectation values. 

After integrating out $\Phi$, we obtain a tree level superpotential which is quadratic in 
terms of the meson field $X$. To get the full effective superpotential, we need to 
include the non-perturbative Affleck-Dine-Seiberg potential \cite{affdisei}. We then get an 
effective superpotential for the meson field of the form 
\begin{equation}
  W_{eff} = (N_c-N_f)\biggl(\frac{\hat\Lambda^{3N_c-N_f}}{\det X}\biggr)^{\frac{1}{N_c-N_f}} + \Tr mX-\frac{1}{2g}\Tr X^2. 
\end{equation}
The equation for a critical point of this superpotential is 
\begin{equation}
   \biggl(\frac{\hat\Lambda^{3N_c-N_f}}{\det X}\biggr)^{\frac{1}{N_c-N_f}}1-mX+\frac{1}{g}X^2 = 0.
\end{equation}
At the critical point we can write the superpotential in a slightly 
simpler form,  
\begin{equation}\label{wcritmeson}
  W_{eff,crit} = \frac{2N_c-N_f}{2}\biggl(\frac{\hat\Lambda^{3N_c-N_f}}{\det X}\biggr)^{\frac{1}{N_c-N_f}}
  +\frac{1}{2}\Tr mX. 
\end{equation}

More generally, we can diagonalize the meson matrix $X$. We can actually assume 
that $X$ is diagonalized simultaneously with the mass matrix $m$. This follows 
from the equations of motion for the flavors, 
\begin{equation}
  0 = \frac{\partial S}{\partial Q}Q-\tilde Q\frac{\partial S}{\partial\tilde Q} = mX-Xm. 
\end{equation}
Note that these equations of motions are exact quantum mechanically, as the 
flavors appear only quadratic in the action. We will therefore call the 
eigenvalues of the meson field $X_a$. 
The equations for the critical points for the eigenvalues can now be written 
\begin{equation}
  \sigma-m_aX_a+\frac{1}{g}X_a^2=0,
\end{equation}
where we have introduced 
$\sigma =  \biggl(\frac{\hat\Lambda^{3N_c-N_f}}{\det X}\biggr)^{\frac{1}{N_c-N_f}}$. 
We can now solve for the eigenvalues of the meson field in terms of 
$\sigma$, finding 
\begin{equation}
  X_a = \frac{\sigma}{m_a\gamma_{\epsilon_a}(\alpha_a\sigma)},
\end{equation}
where $\epsilon_a=\pm$. 
Due to the obvious relation $\det X = \prod_aX_a$, 
$\sigma$ has to satisfy the relation 
\begin{equation}
  \prod_{a=1}^{N_f}\gamma_{\epsilon_a}(\alpha_a\sigma)
  = \frac{\sigma^{N_c}}{\hat\Lambda^{3N_c-N_f}\det m}.
\end{equation}
We now find that if we relate the scales as 
$\hat\Lambda^{3N_c-N_f} = \frac{\Lambda^{3N_c}}{\det m}$, 
and we replace $\sigma$ by $S$, this is precisely the equation \eqref{scrit} 
for $S$. Moreover with this identification also the effective superpotentials 
\eqref{wcrits} and \eqref{wcritmeson} agree. 

As the choice of $\epsilon_a$ is related to the choice of sheet on which the 
point corresponding to $m_a$, we see that the two different branch cuts 
in the $A_2$ model near $x=m_a$ correspond to the choice of eigenvalues 
for the meson field.

\section{Conclusions and Discussion}
\label{sec:concl}

We studied effective glueball superpotentials for $\CN=1$ super Yang-Mills theories 
with fundamental matter and an adjoint Higgs field. We calculated these using 
matrix models at large $N$ with a large number of flavors. In particular, we
used the 2-matrix model in a special limit. As noted earlier, this 
allows us to deal exactly, and purely in a large $N$ limit, with arbitrary 
tree level superpotentials for the adjoint field. Although we only 
worked out in detail the Gaussian case, more general superpotentials 
are straightforwardly added, but technically more difficult to solve exactly. 

The description of flavors using the $A_2$ quiver matrix model has some intriguing 
properties. The presence of two types of cuts in the $A_2$ model 
connecting the sheet corresponding to the flavor 
As noted above, we can see a sign of Seiberg-like duality from the 
$\Z_2$ action interchanging two sheets. It would be interesting to 
pursue this further.

\section*{Acknowledgements}

We would like to thank Christian R\"omelsberger, Alessandro Tomasiello, Sujay Ashok, 
and Michael Douglas for useful discussions. 
This research was partly supported by DOE grant \#DE-FG02-96ER40959.

\end{document}